\documentclass[a4paper]{jpconf}
\usepackage{xspace}
\usepackage{graphicx}
\usepackage{units}
\newcommand{\wz}{\texttt{WHIZARD}\xspace}
\newcommand{\om}{\texttt{O'Mega}\xspace}
\newcommand{\vamp}{\texttt{VAMP}\xspace}
\newcommand{\circe}{\texttt{CIRCE}\xspace}
\newcommand{\go}{\texttt{GoSam}\xspace}
\newcommand{\ol}{\texttt{OpenLoops}\xspace}
\newcommand{\si}{\texttt{SINDARIN}\xspace}
\newcommand{\toppik}{\texttt{TOPPIK}\xspace}

\usepackage[hyphens]{url}
\usepackage[babel]{csquotes}
\usepackage[style=numeric,
  maxnames=3,
  minnames=3,
  backend=bibtex,
  citestyle=numeric-comp,
  url=false,
  maxbibnames=5
  ]{biblatex}
\renewbibmacro{in:}{}
\bibliography{ACAT-Proceedings}
\AtEveryBibitem{\clearfield{month}}
\AtEveryBibitem{\clearfield{title}}
\begin{document}
\begin{flushright}
	DESY 16-033, UWTHPH 2016, MITP/16-022, SI-HEP-2016-XX, LTH 1077
\end{flushright}
\vspace{-5em}
\title{Automation of NLO processes and decays and POWHEG matching in \wz}
\author{J\"urgen Reuter$^1$, Bijan Chokouf\'{e}$^1$, Andr\'{e} Hoang$^{2,3}$, Wolfgang Kilian$^4$, 
Maximilian Stahlhofen$^{5,1}$, Thomas Teubner$^6$, Christian Weiss$^{1,4}$}
\address{$^1$ DESY Theory Group, Notkestr. 85, 22607 Hamburg, Germany}
\address{$^2$ University of Vienna, Faculty of Physics, Boltzmanngasse 5, 1090 Vienna, Austria}
\address{$^3$ Erwin Schr\"odinger International Institute for Mathematical Physics, University of Vienna, 
    Boltzmanngasse 9, 1090 Vienna, Austria}
\address{$^4$ University of Siegen, Emmy-Noether Campus, Walter-Flex-Str. 3, 57068 Siegen, Germany}
\address{$^5$ PRISMA Cluster of Excellence, Institute of Physics, Johannes Gutenberg University, 
    55128 Mainz, Germany}
\address{$^6$ University of Liverpool, Department of Mathematical Sciences, Liverpool L69 3BX, U.K.}
\ead{juergen.reuter@desy.de, bijan.chokoufe@desy.de, andre.hoang@univie.ac.at, kilian@physik.uni-siegen.de
     mastahlh@uni-mainz.de, thomas.teubner@liverpool.ac.uk, christian.weiss@desy.de}

\begin{abstract}
        We give a status report on the automation of next-to-leading
        order processes within the Monte Carlo  
        event generator WHIZARD, using GoSam and OpenLoops as provider
        for one-loop matrix elements.  
        To deal with divergences, WHIZARD uses automated FKS
        subtraction, and the phase space for singular regions  
        is generated automatically. NLO examples for both scattering
        and decay processes with a focus on $e^+e^-$  
        processes are shown. Also, first NLO-studies of observables
        for collisions of polarized leptons beams,  
        e.g. at the ILC, will be presented. Furthermore, the automatic
        matching of the fixed-order NLO amplitudes  
        with emissions from the parton shower within the POWHEG
        formalism inside WHIZARD will be discussed.  
        We also present results for top pairs at threshold in lepton
        collisions, including matching between a  
        resummed threshold calculation and fixed-order NLO. This allows the investigation of 
        more exclusive differential observables.
\end{abstract}
\section{The \wz Event Generator}
\wz \cite{kilian07wh} is a multi-purpose event generator for in principle arbitrary processes at hadron and lepton colliders. Being able to treat beam-spectra and initial-state photon radiation, it is especially suited for lepton collider physics studies. Moreover, its built-in scripting language \si allows for the analysis in the same framework.
\wz is written in Fortran2003. Its structure is strictly object-oriented, so that a modular setup enables
the convenient interface to numerous other programs. The main sub-components of \wz are \om\cite{moretti01om}, \vamp\cite{ohl1999va} and \circe\cite{ohl1997ci}:
\om provides multi-leg tree-level matrix elements using the helicity formalism. \vamp is used for Monte-Carlo integration and grid sampling. \circe creates and evaluates lepton beam spectra. Color factors are evaluated using the color-flow formalism \cite{kilian12co}.
\wz can be used for event generation on parton level as well as for the subsequent shower simulation. For this purpose, it has its own analytical and $k_T$-ordered parton shower \cite{kilian12an} as well as a built-in interface to \texttt{Pythia6} \cite{Sjoestrand2006py}. An automated interface to \texttt{Pythia8} \cite{Sjoestrand2015py} or \texttt{HERWIG++} \cite{bahr2008he} is not yet present, but planned.
Though \wz spearheaded many BSM phenomenological studies \cite{kilian2005lh, hagiwara2006su, 
beyer2006el, alboteanu2008re, kalinowski2008sn, kilian2015hi, kilian2015res},
these proceedings focus on the automation of SM QCD NLO corrections and non-relativistic top threshold resummation.
\section{Automated NLO Calculations in \wz}
Next-to-leading order (NLO) calculations have become standard for the prediction of most observables 
during the last decade. Substantial progress has been made in the automation of loop matrix-element 
computations as well as in NLO parton shower matching \cite{nason2004po, frixione2007po}. 
At the LHC, NLO simulations are routinely employed. Computer programs in this field cover a range from dedicated, 
single-purpose codes to automated multi-purpose event generators.\\
With the ILC approaching its possible approval phase, ILC studies become more specific \cite{baer2013ilc}, 
thereby increasing the need for easy to use NLO programs. Here, we want to combine 
the expertise of \wz in the field of lepton collisions with the improved accuracy of NLO predictions. There
have been earlier works on NLO QED extensions to \wz for certain supersymmetric processes \cite{kilian2006nlo, 
kalinowski2008nlo} as well as on NLO QCD corrections for $pp \to b\bar{b}b\bar{b}$ \cite{binoth2010nlo,
greiner2011nlo} using Catani-Seymour subtraction. However, the fully generic NLO framework is a recent development.\\
Automated NLO programs deal with the infrared divergences of real and virtual matrix elements by
using subtraction procedures. These rely on the extraction of the soft and collinear limit in its
most generic form. This is possible because the $\mathcal{O}(\alpha_s)$-corrections factorize
in the corresponding limits. These subtraction terms $d\sigma^{\rm{S}}$ are then added and subtracted, such that
\begin{equation}
	d\sigma^{\rm{NLO}} = d\sigma^{\rm{LO}} 
            + \underbrace{\int_{n+1} \left (d\sigma^{\rm{R}} - d\sigma^{\rm{S}}\right)}_{\rm{finite}} 
            + \underbrace{\int_{n+1} d\sigma^{\rm{S}} + \int_n d\sigma^{\rm{V}}}_{\rm{finite}}.
	\label{Eqn:soft_mismatch}
\end{equation}
The explicit form of $d\sigma^{\rm{S}}$ is arbitrary. The two most common schemes employed are
Catani-Seymour subtraction \cite{catani97cs} and the FKS (Frixione-Kunszt-Signer) scheme \cite{frederix2009fks}. \wz uses the latter. 
The program automatically finds the singular regions involved in the desired process and computes
the subtraction terms to the real matrix element correspondingly.\\
NLO computations require virtual loop matrix elements. However, \om can only generate tree-level matrix
elements. So, to obtain virtual matrix elements, external programs, so-called One-Loop Providers (OLP),
can be interfaced to \wz. Up to now, there are working interfaces to \go \cite{Cullen2012go} and \ol \cite{Cascioli2012ol}.
They use the Binoth Les Houches Accord (BLHA) \cite{alioli2014blha}, which unifies the interface between Monte-Carlo generator and OLPs. 
The BLHA interface in \wz is generic, meaning that every other matrix element generator which complies to the BLHA
standard can be interfaced very easily in the same fashion.\\
The NLO functionality of \wz has first been presented
in~\cite{Weiss2015nlo}, where we have focussed 
our discussion on fixed-order off-shell top production. In the following, we want to present the
newest developments in \wz for NLO.

\subsection{Treatment of Resonances}

A recent development in \wz is the special treatment of radiative
corrections occuring from 
particles which originate from a common resonance. This issue has been
extensively discussed in~\cite{jezo2015re}. 
We have automated the scheme discussed therein in \wz.

Subtraction methods like FKS lose accuracy when after the
construction of the real phase space, the resonance momentum deviates
significantly from its value 
in the Born phase space. This affects the propagator both in the real
and the Born matrix element, 
which is part of the subtraction terms. The real matrix element and
its approximation therefore 
do not match perfectly, impairing the convergence of the integration
and distorting the invariant 
mass spectrum in event generation \cite{jezo2015re}.

\begin{figure}
    	\centering
	\begin{minipage}{22pc}
		\includegraphics[width=22pc]{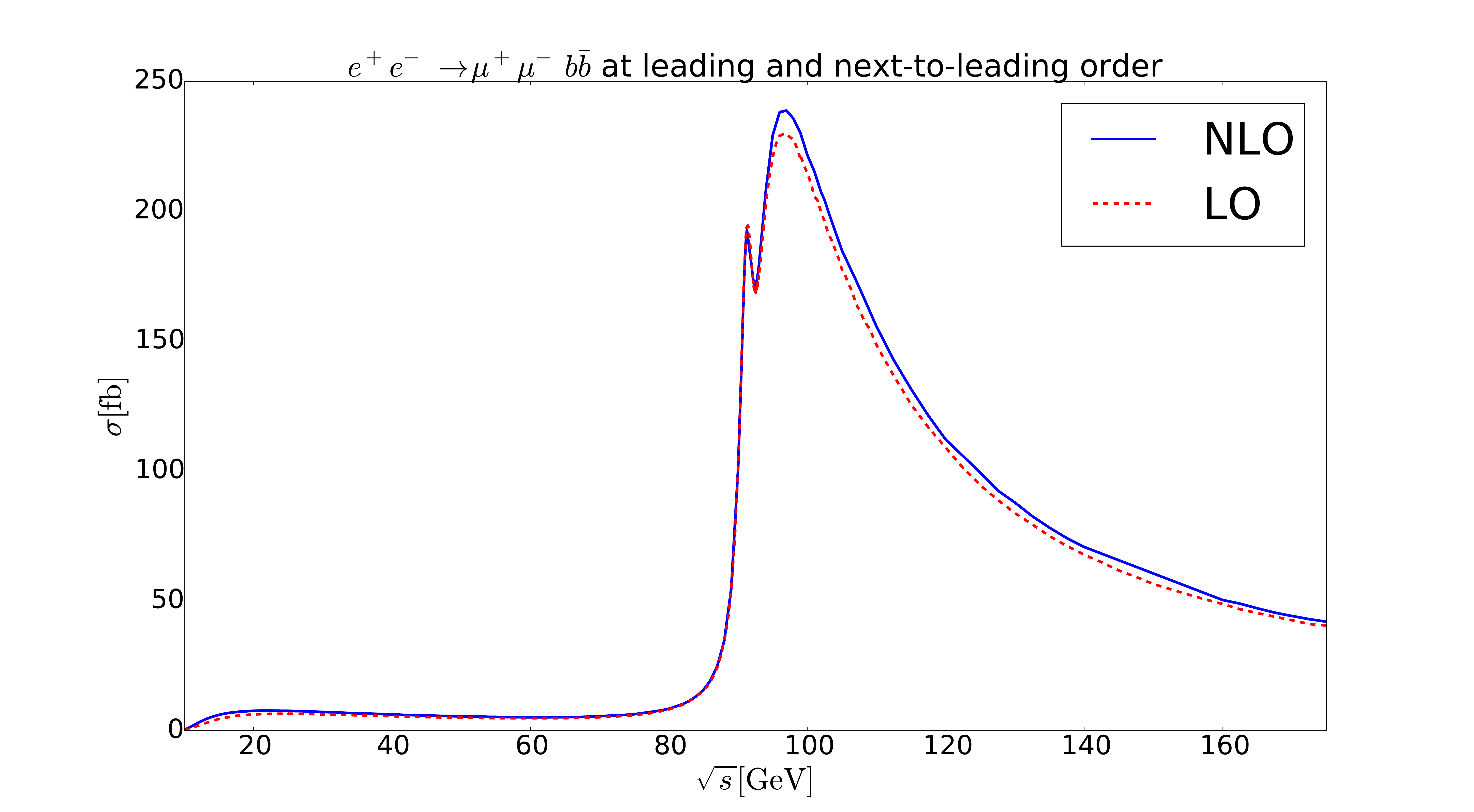}
	\end{minipage}
	\begin{minipage}{17pc}
		\includegraphics[width=17pc]{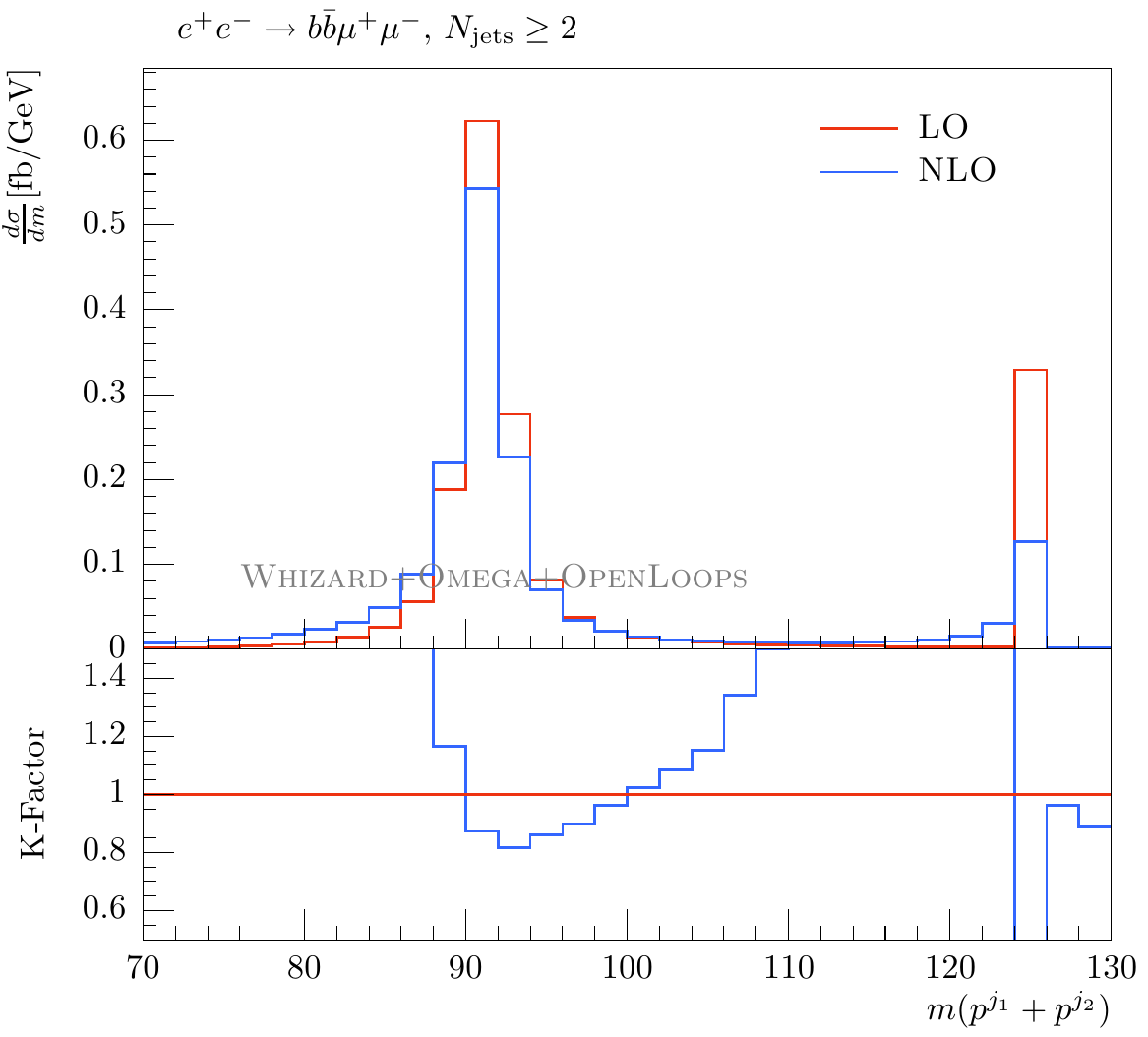}
	\end{minipage}
	\caption{\label{fig:resonance_plots} Left: $\sigma_{\rm{tot}}$ at the $Z\mu\mu$ and $Zbb$ resonance. 
		Right: Invariant mass distribution of the two hardest jets at $\sqrt{s} = \unit[800]{GeV}$. 
		Jets have been clustered using \texttt{FastJet} \cite{Cacciari2014fj}. 
                The $Z$ and especially the $H$ peak are being washed out by gluon emissions.}
\end{figure}

This problem is solved by employing a phase space mapping that
generates the real phase space 
in such a way that the resonance momentum is conserved. Thus the real
matrix element is evaluated 
at the same point on the resonance's Breit-Wigner curve. However,
this approach comes with the drawback of increased complexity in the
implementation, since now 
the calculation has to be aware of the additional singular regions
belonging to different 
resonance histories. Moreover, the so-called soft mismatch (Eq.~(3.59)
in \cite{jezo2015re}) 
enters the calculation as an additional new component of the integration. This is due to the fact
that the sum over all singular regions does not reproduce the full real matrix element in this
approach, because for each singular region, the FKS mapping is evaluated in the associated
resonance's frame of reference. However, usual FKS requires all subtraction terms to be evaluated in the
same frame of reference. This property can be restored, yielding the soft mismatch term.\\
\begin{table}
	\caption{Comparison of integration histories for the standard
          and resonance-aware subtraction  
          at $\sqrt{s} = \unit[800]{GeV}$. The Coulomb
          singularity which arises from $\gamma \rightarrow
          \mu^+ \mu^-$ 
          has been cut out by imposing $m_{\mu\mu} > \unit[20]{GeV}$. Each
          iteration uses 12000 calls.} 
	\begin{center}
		\begin{tabular}{lllll}
			\br
			$n_{it}$ & $\sigma[fb]$, standard & Error[\%] & $\sigma[fb]$, resonances & Error[\%] \\
			\mr
			1 & 9.68118 & 66.30 & 2.90570 & 2.87 \\
			2 & 2.85397 &  8.25 & 2.85920 & 1.82 \\
			3 & 2.49076 & 26.25 & 2.92779 & 1.40 \\
			4 & 2.76956 & 34.91 & 2.85123 & 1.40 \\
			5 & 2.43462 & 19.80 & 2.88554 & 1.34 \\
			\br
			5 & 2.75391 &  7.15 & 2.88420 & 0.71 \\
			\br
		\end{tabular}
	\end{center}
	\label{Table:comparison}
\end{table}
\begin{table}
	\caption{Real-subtracted integration component and, in the case of resonance-aware sutbraction,
	soft mismatch, for $\Gamma_H = \unit[1000]{GeV}$ at $\sqrt{s} = \unit[800]{GeV}$.}
	\begin{center}
		\begin{tabular}{llll}
			\br
			Method & $\sigma_{\rm{soft}}[fb]$ & $\sigma_{\rm{mism}}[fb]$ & $n_{\rm{calls}}$ \\
			\mr
			standard & -3.31997 $\pm$ 0.62\% & 0 & 5 $\times$ 100000 \\
			resonances & -1.62098 $\pm$ 0.29\% & -1.70388 $\pm$ 0.56\% & 
			5 $\times$ 20000 (soft) + 5 $\times$ 20000 (mismatch) \\
			\br
		\end{tabular}
	\end{center}
	\label{Table:RealConvergence}
\end{table}
The implementation of the resonance-aware subtraction has been applied to the process 
$e^+ e^- \rightarrow b \bar{b} \mu^+ \mu^-$, where there is one resonance topology with two associated
resonance histories, namely $Z/H \rightarrow b\bar{b}$. This process is also of interest in studies of 
Higgsstrahlung processes at lepton colliders. 
We have set $m_b = \unit[4.2]{GeV}$, so that collinear divergences do not occur. 
Without resonance mappings, we observe that the integration
does not perform well, due to the very narrow Higgs resonance ($\Gamma_H \sim \unit[4]{MeV}$). Even a relatively soft
gluon emission is enough to alter the resonance momentum in such a way that the soft approximation
does not fit well enough with the real matrix element. Using the resonance-aware subtraction we obtain
significantly better results, as can be clearly seen in Table \ref{Table:comparison}. Taking the large-width limit
yields a good integration also for the non-resonant case, as desired. Table \ref{Table:RealConvergence} shows that both
calculations give the same result in this limit. However, the non-resonant approach needs significantly
more integration calls to reach the same accuracy as its resonance-aware counterpart.
Figure \ref{fig:resonance_plots} shows a scan of the total cross section. 
The generation of fixed-order NLO events has been modified for the use with resonance-aware subtraction. In
the earlier version of \wz, weighted $N$-particle events were produced with weight 
$\mathcal{B} + \mathcal{V} + \mathcal{S}$ and associated with $N+1$-particle events with weight
$\mathcal{R}_{\alpha_r}$ for each singular region $\alpha_r$. The events are saved in \texttt{HepMC}-files.
If resonances are included, different $\alpha_r$
can yield the same phase space if they only differ by the underlying resonance mass. Therefore, to save
disk space, the number of real events now is the number of different possible phase spaces. Their weight is
the sum of all $\mathcal{R}_{\alpha_r}$ whose $\alpha_r$ belong to the same phase space.
Figure \ref{fig:resonance_plots} shows an example of NLO event generation in \wz.
\subsection{Other features}
\wz is able to produce events according to the POWHEG scheme, which can be used as input to parton showers
and preserve the NLO accuracy of the subsequent simulations. This has been discussed in \cite{Nejad2015pow} and will
not be elaborated further here.\\
Apart from the major NLO functionalities above, \wz also can now deal with several minor aspects of NLO
calculations. First, \wz can also compute decay widths at next-to-leading order and generate fixed-order
events accordingly. The first use case here is the top decay $t \rightarrow bW$. The computation takes into account both
the initial-state gluon emissions from the top quark and final-state emissions from the bottom quark.
The results of this computation play a crucial role in the study of the top threshold described in the
next section, where it is important to keep consistency between the input parameters and the top width
used.\\
Second, NLO calculations can be performed using polarized lepton beams. This is a desirable feature for future
studies at the ILC or CLIC, which will operate with polarized leptons. For this purpose, \wz uses a 
modified BLHA interface to \ol, which allows it to provide the virtual matrix elements individually
for each helicity configuration of the colliding lepton pair.
\section{Top Threshold Resummation and Matching to the Continuum}
\begin{figure}[htb]
	\centering
	\includegraphics[width=30pc]{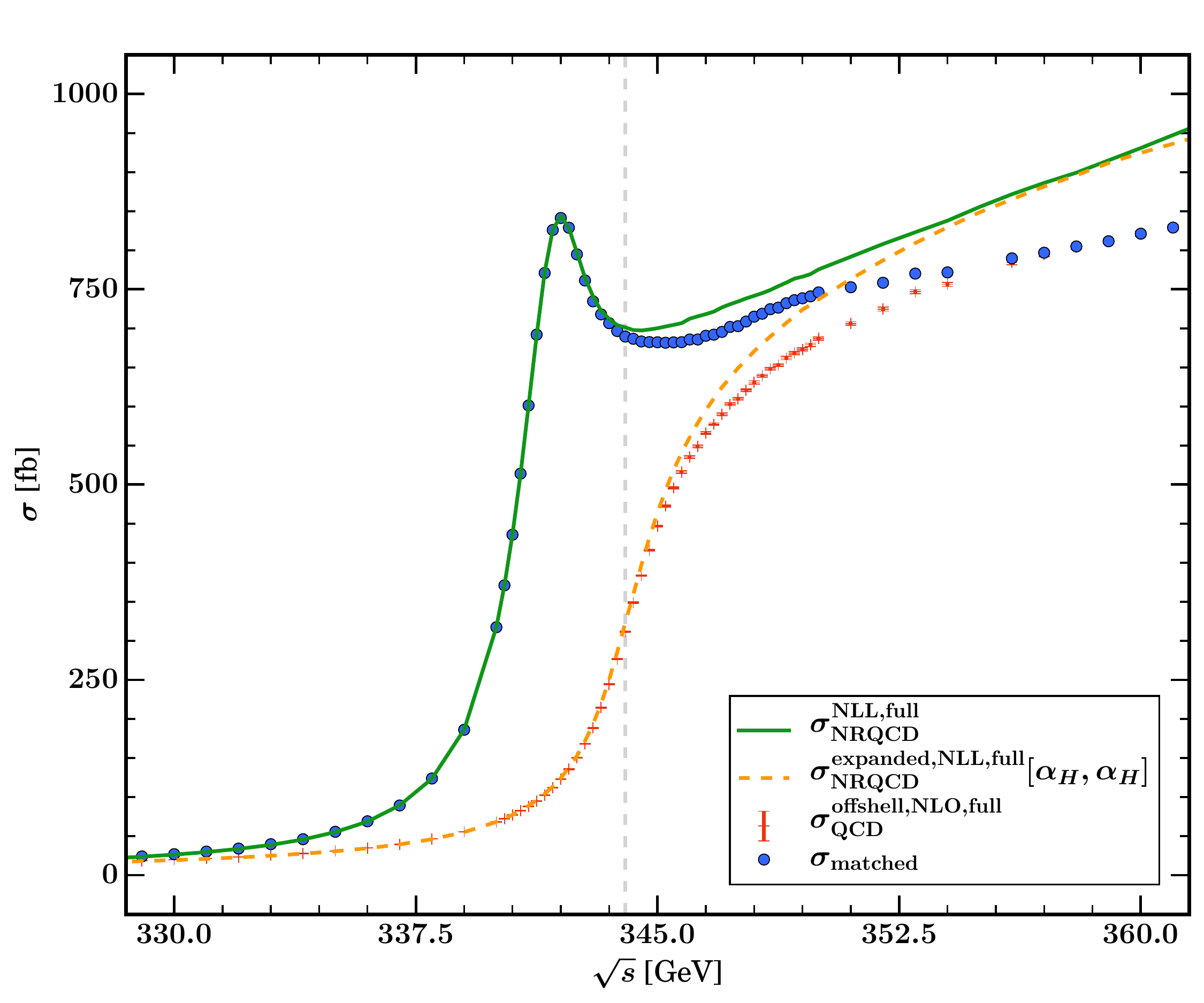}
	\caption{\label{fig:matching} Matching the NLL resummed threshold prediction to the fixed-order NLO QCD
	         continuum for the total cross section of the process $e^+ e^- \rightarrow W^+bW^-b$. The blue dots
		 show the result obtained by the (preliminary) matching prescription between vNRQCD and the continuum.
	         The solid green line corresponds to the insertion  of the
	         NLL form factor from \toppik into the Born process. The dashed orange curve shows the same 
	 	 result with the form factor expanded to first order in $\alpha_s$. The red crosses represent the
		 full relativistic fixed-order NLO result.}
\end{figure}
High-energy lepton colliders operating at the top threshold or close
to it allow for the most precise method 
to measure the mass of the top quark known to date. However, at
threshold the top quark pair is non-relativistic 
and can thus form a toponium quasi-bound state, due to the attractive
top quark potential. This potential relies 
on the exchange of an arbritrary number of virtual gluons between the
tops. The bound-state effects give rise to 
Coulomb singularities $(\alpha_s / v)^n$ and large logarithms $\ln^n
v$ in the perturbative expansion 
of the cross section. Here, $v \sim \alpha_s \sim 0.1$ is the relative
velocity of the non-relativistic top quarks. 
The resummation of the threshold enhanced terms can be carried out in
the vNRQCD framework \cite{luke2000nrqcd, Pineda2002, hoang2006hq, hoang2010top,
  hoang2011nll}. The R-ratio then takes the schematic form
\cite{hoang2014nnll} 
\begin{equation}
	R = \frac{\sigma_{t\bar{t}}}{\sigma_{\mu\mu}} = v \sum_k \left(\frac{\alpha_s}{v}\right)^k 
		\sum_l \left(\alpha_s \ln v\right)^l \times
		\left\lbrace \underbrace{1(LL); \alpha_s, v (NLL);}_{\rm{effective\ ttV\ vertex}} \alpha_s^2, \alpha_s v, v^2 (NNLL); ...\right\rbrace.
\end{equation}
In \wz, the resummation is achieved by replacing the $tt\gamma$ and
$ttZ$ vertices by non-relativistic form factors 
at NLL accuracy using the external code \toppik
\cite{Hoang1999to}. This interface was first discussed in
\cite{bach2015wh}. 
A recent development is the matching of the vNRQCD-approximation to
the relativistic continuum, complementing the 
accuracy of the two approximations. 
The matching between the resummed prediction at threshold and the
relativistic NLO result, which is the reliable result 
for $\sqrt{s} \gg 2m_t$, is based on two main concepts. First, we have
to subtract the first order expansion of the 
resummed computation when we want to add the NLO contributions. This
has to be done in a way that respects the relevant 
scales of the vNRQCD calculation (hard $m_t$, soft $m_t v$ and
ultra-soft $m_t v^2$). In Figure \ref{fig:matching} we can 
see how this expansion evaluated at the hard scale ($\alpha_H =
\alpha_s{m_t}$), shown as orange dashes, reproduces the full NLO
result,  
represented by red crosses, as it contains the dominant terms close to
threshold. On the other hand, we face the problem 
that the resummed prediction keeps growing arbitrarily with
$\sqrt{s}$, which is an artifact of the assumption that 
the computation is performed close to threshold and is seen by the
rise of the green curve in Fig. \ref{fig:matching}. 
This is cured by multiplying the relevant scales with a switch-off
function that smoothly approaches zero  
as one moves away from threshold.
Combining these concepts gives a nice physical prediction in form of
the blue dots. Note that while we have concentrated 
on the inclusive cross section here, the implementation in \wz will
eventually allow for the  
first time for abritrary differential distributions with NLO+NLL
precision together with their uncertainties based on 
scale variation.

\section{Acknowledgements}
We want to thank Jonas Lindert for many useful discussions and
especially the fast \ol support. We also thank Tom\'{a}\v{s} Je\v{z}o
for his advice about the resonance-aware FKS subtraction. JRR wants to
thank the ACAT 2016 organizers for the beautiful venue in
Valpara\'{i}so and a fantastic conference in Chile. 
\printbibliography
\end{document}